\documentclass[12pt,preprint]{aastex}

\begin{document}

\title{Discovery of a W UMa type binary GSC 03553-00845}

\author{Guo, D.-F.\altaffilmark{1,2}, Li Kai\altaffilmark{1,2,3}, Hu, S.-M.\altaffilmark{1,2}, Jiang, Y.-G.\altaffilmark{1,2}, Gao, D.-Y.\altaffilmark{1,2}, Chen, X.\altaffilmark{1,2}}

\altaffiltext{1}{Institute of Space Sciences, School of Space Science and Physics, Shandong University, Weihai, 264209, China (e-mail: kaili@sdu.edu.cn, likai@ynao.ac.cn (Li, K.); husm@sdu.edu.cn (Hu, S.-M.))}
\altaffiltext{2}{Shandong Provincial Key Laboratory of Optical Astronomy and Solar-Terrestrial Environment, Weihai, 264209, China}
\altaffiltext{3}{Key Laboratory for the Structure and Evolution of Celestial Objects, Chinese Academy of Sciences}

\begin{abstract}
When observing the transiting extrasolar planets, we discovered a new W UMa type binary GSC 03553-00845. Following observation was carried out using the 1m telescope at Weihai Observatory of Shandong University. Complete BVR light curves were determined. Using the W-D program, we analyzed the light curves. Photometric solution reveals that GSC 03553-00845 is a W-subtype W UMa type binary with a mass ratio of $q=2.904$, it is an overcontact binary system by a contact degree of $f=29.5\%$ with a small temperature difference between the components ($\Delta T=206$ K) indicating a good thermal contact between the components. More observation of GSC 03553-00845 is needed in order to analyze the light curve variation and orbital period change.
\end{abstract}


\section{Introduction}

W UMa type stars are usually contact binaries with two low temperature components with spectral types of F to K. Both component stars are in contact with each other and sharing a common convective envelope. They have two typical characteristics. These are (1) the effective temperatures of both components are very similar, and (2) the currently less massive component is overluminous for its current mass. The W UMa systems are frequently classified into two subtypes: A-types and W-types, which is according to whether the larger or smaller star has the higher temperature (Binnendijk 1970). In A-types, the more massive star is eclipsed at primary minimum and is the hotter component, while the less massive star is eclipsed at primary minimum for the W-types (i.e., the more massive component is the cooler one). Although much is known about W UMa binaries (e.g., Liu et al. 2011; Qian et al. 2013; Liao et al. 2014), a complete theory of their origin, structure, and evolution has not been developed. They are thought to form as detached binaries and ultimately decay into contact binaries by magnetic braking.

In this paper, we presented the investigation of multiple color light curves of a newly discovered W UMa type binary named GSC 03553-00845 ($\alpha_{2000}=18^{h}57^{m}15.4^{s}$, $\delta_{2000}=+51^{\circ}16^{\prime}31.9^{\prime\prime}$). This study follows the following structure. In Section 2, we present the discovery, CCD observations and the orbital period determination of GSC 03553-00845. Photometric analysis of the eclipsing pair is shown is Section 3. In Section 4, the results are summarized.

\section{Observations, discovery and period determination}
While conducting an observing program directed toward the study of transiting extrasolar planets (HAT-P-37b), it was noticed that one of the comparison stars varied by more than 0.5 magnitudes (V band) on the nights of September 18, October 2 and October 18, 2013. Since this star is not listed in neither the GCVS nor the NSV catalogues, we conclude that its variability has not been previously reported. The new variable star was identified to be GSC 03553-00845 according to the GSC 1.2.

Additional differential BVR photometry was obtained using a PIXIS 2048B CCD camera attached to the 1.0 m Cassegrain telescope (Hu et al., 2014) at Weihai Observatory of Shandong University on the nights of April 29, 30 and May 18, 2014. The PIXIS camera has $2048\times2048$ square pixels and the pixel size is 13.5$\mu$m. The scale of the image is about 0.35$\arcsec$ per pixel and the effective field of view is about 11.8$'$ $\times$ 11.8$'$. The standard Johnson and Cousins filters ($B$, $V$, and $R$) were used during our observations. The typical exposure time for each image was 40 s in B band, 20 s in V band and 10s in R band. The reductions of observations were conducted using the IMRED and APPHOT packages in IRAF\nolinebreak\footnotemark[1] procedures.\footnotetext[1]{IRAF is distributed by the National Optical Astronomy Observatories, which is operated by the Association of Universities for Research in Astronomy Inc., under contract to the National Science Foundation.} All data were processed by bias and flat-field correction. One of the CCD images is shown in Figure 1, where "V" refers to the variable star (i.e., GSC 03553-00845), "C" to the comparison
star, and "CH" to the check star. The comparison star is GSC 03553-00840 ($\alpha_{2000.0}=18^h57^m11^s.7$, $\delta_{2000.0}=51^{\circ} 15^{\prime}41$\arcsec$ .05$)  and check star is GSC 03553-00567 ($\alpha_{2000.0}=18^h57^m26^s.2$, $\delta_{2000.0}=51^{\circ} 17^{\prime}20$\arcsec$ .43$).

Jurkevich method (Jurkevich 1971) was applied to all the V band data for periodicity analysis. Jurkevich method is based on the expected mean square
deviation. It does not require an equally spaced observations, so it is less
inclined to generate a spurious periodicity compared to a Fourier analysis. 
Jurkevich method tests a series of trial periods and the data are
folded according to the trial periods. Then all data are divided
into $m$ groups according to their phases around each trial period. The variance $V_{i}^{2}$
for each group and the sum of each group variance $V_{m}^{2}$  are
computed. If a trial period equals to the real one, then $V_{m}^{2}$
would reach its minimum. The detailed computation of the variances
is described in Jurkevich (1971). The possible periods can be easily obtained from the plot. The results derived by the Jurkevich method with  m=60 are shown in Figure 2. The minimum value $V^{2}_{m}=0.013$ indicates the period of 0.435470 days.

\begin{figure*}
\begin{center}
\includegraphics[angle=0,scale=0.7]{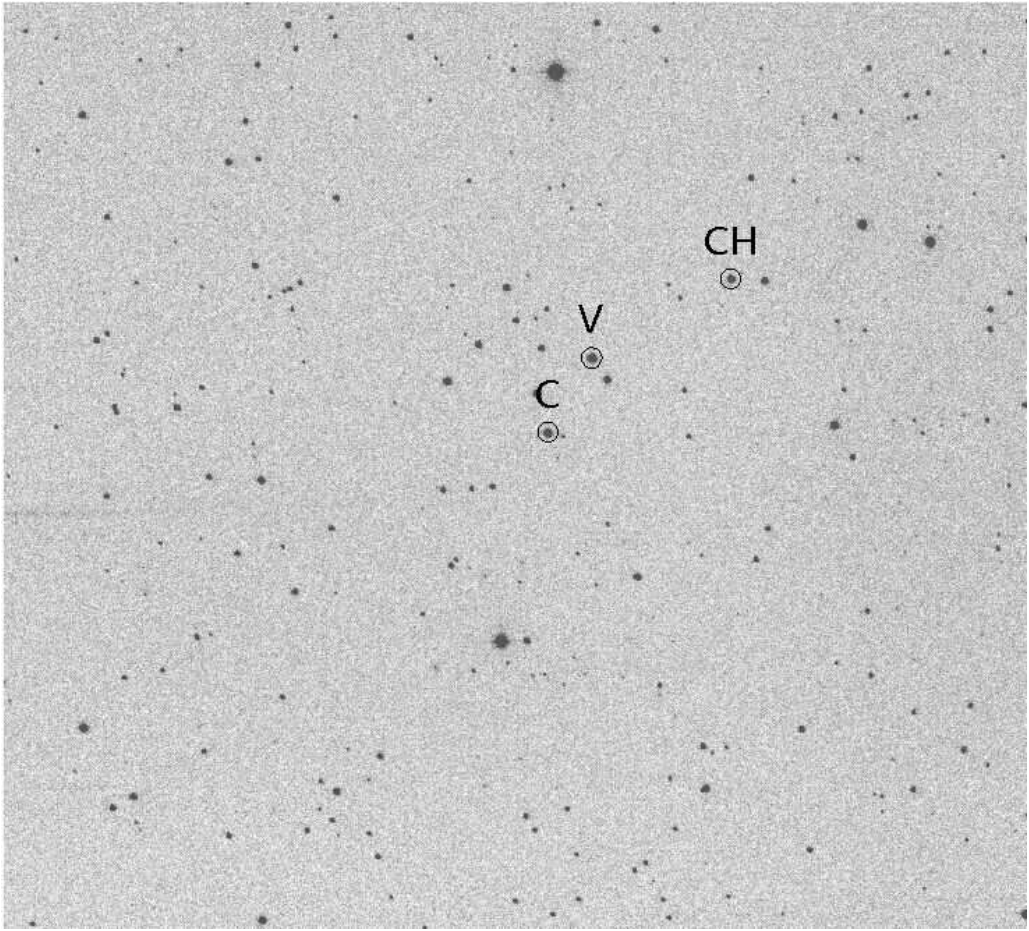}

\caption{CCD image in the field of view around GSC 03553-00845. "V" refers to the variable star (i.e., GSC 03553-00845), "C" refers to the
comparison star, while "CH" refers to the check star.}
\end{center}
\end{figure*}

\begin{figure*}
\begin{center}
\includegraphics{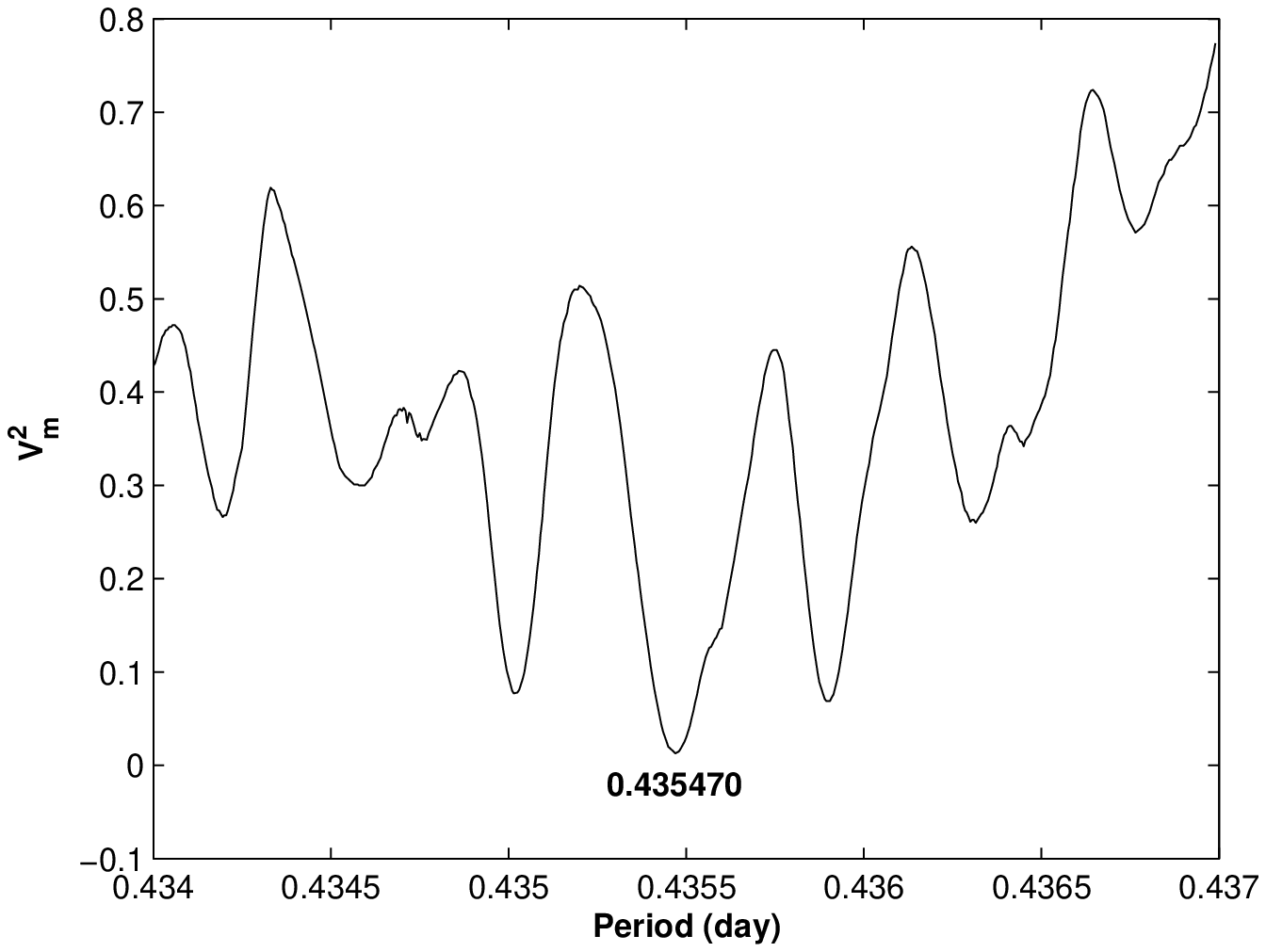}

\caption{Relationship between the trial period and $V_{m}^{2}$ using all the V band data.}
\end{center}
\end{figure*}

The V band light curve phased with the period 0.435470 d is shown in Figure 3. Additional differential BVR light curves are shown in Figure 4. From the two figures, it is seen that the data
during six days merged smoothly and the light variation is of W UMa
type eclipsing binary. The curves show some scatter which is not unusual for WU Ma binaries which show night to night variations
due to intrinsic activity. However, some of this variability may be due to weather conditions at the observing site. Four times of light minimum were determined, they are: HJD2456554.0814$\pm$0.0003, HJD2456568.0175$\pm$0.0002, HJD2456777.2574$\pm$0.0003, and HJD2456796.1997$\pm$0.0009.

\begin{figure*}
\begin{center}
\includegraphics{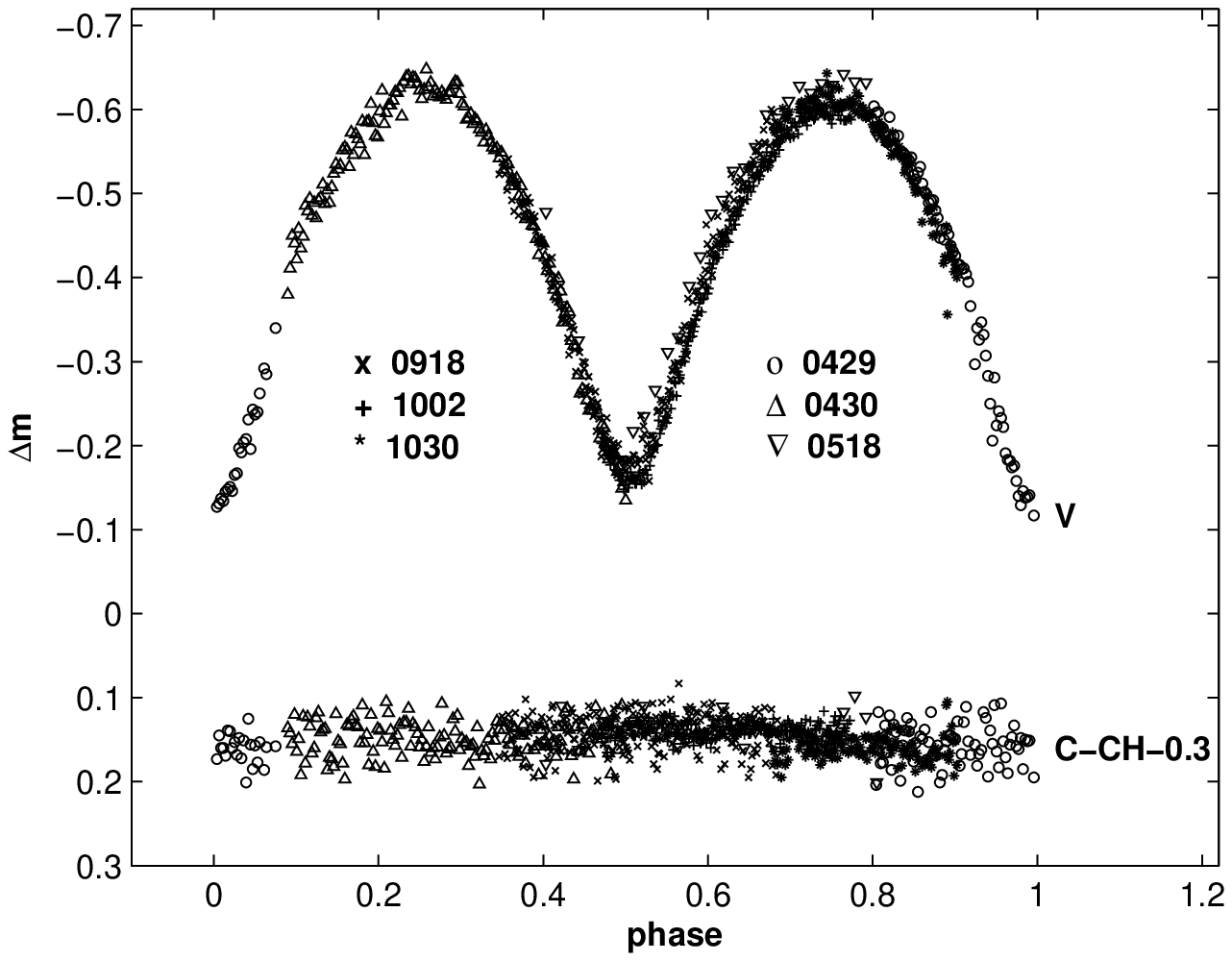}

\caption{Observed V light curve of GSC 03553-00845 based on all data. The phases were calculated using the
equation Min.I = HJD2456777.2574 + 0.435470E. Different symbols represent
different days.}
\end{center}
\end{figure*}

\begin{figure*}
\begin{center}
\includegraphics{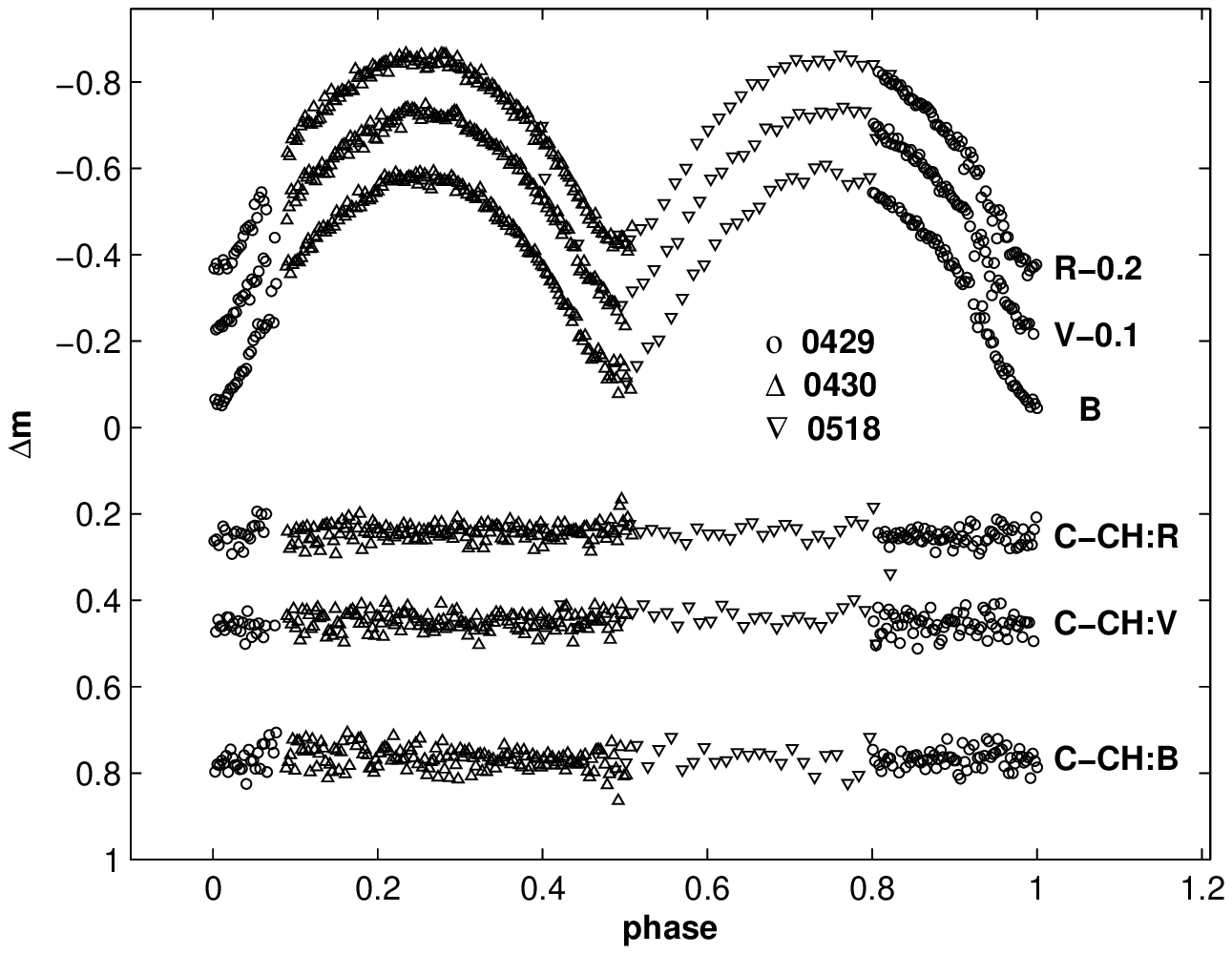}

\caption{BVR light curves of GSC 03553-00845 observed on April 29, 30 and May 18, 2014. The phases were calculated using the
equation Min.I = HJD2456777.2574 + 0.435470E. Different symbols represent
different days.}
\end{center}
\end{figure*}

\section{Modeling the light curves}

The photometric solutions of GSC 03553-00845 were obtained using the W-D program (Wilson \& Devinney 1971; Wilson 1990, 1994).
According to NOMAD (The Naval Observatory Merged Astrometric Dataset, Zacharias et al. 2004), the color index of GSC 03553-00845 can be determined as $B-V=0.53$, which is corresponding to the spectral type of F8 due to Cox (2000). Therefore, we fixed the effective temperature of the primary component, $T_1$, at 6250 K. The bolometric and bandpass limb-darkening coefficients were fixed from van Hamme (1993). The bolometric albedo and gravity-darkening coefficient were assumed to be $A_{1,2}=0.5$ and $g_{1,2}=0.32$ for both components.The solutions were started with mode 2, but quickly converged when both components overflowed their Roche lobes. So the iterations were finally made in mode 3. During the solutions, the adjustable parameters were the mass ratio $q$, the effective temperature of the secondary component $T_2$, the monochromatic luminosity of primary component in each band $L_1$, the orbital inclination $i$ and the dimensionless potential of the two component $\Omega_1=\Omega_2$.

Since GSC 03553-00845 is a newly discovered binary, no mass ratio has been determined, a $q$-search method was used to determine its mass ratio. The solutions for several assumed values of the mass ratio $q$ were obtained. The weighted sum of the squared residuals $\sum W_i(O-C)_i^2$ for all the assumed values of $q$ are shown in Figure 5. As displayed in Figure 5, the minimum value of the weighted sum of the squared residuals $\sum$ is around $q=2.90$. Then, we chose $q=2.90$ as the initial value and started a differential correction so that the iteration converges by setting $q$ as a free parameter. The determined results are listed in Table 1. As GSC 03553-00845 is not a totally eclipsing binary system, the $q$-search is not strongly determined, radial velocity curves are needed to confirm the mass ratio given here. The comparison between observed and the theoretical light curves is shown in Figure 6.

\begin{figure*}
\begin{center}
\includegraphics{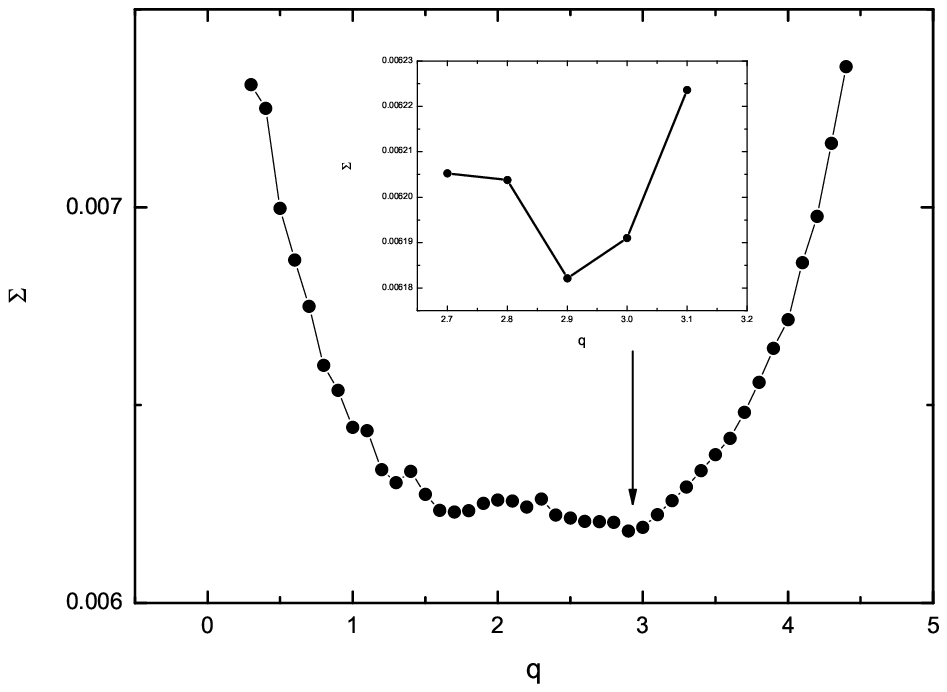}
\caption{$\sum-q$ curves for GSC 03553-00845, the small insert figure is an enlargement that $q$ is from 2.70 to 3.10.}
\end{center}
\end{figure*}

\begin{figure*}
\begin{center}
\includegraphics{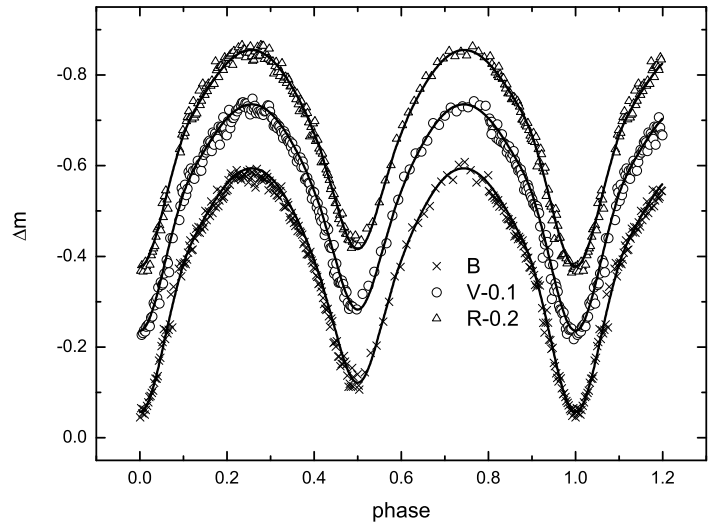}
\caption{The comparison between observed and the theoretical light curves for GSC 03553-00845, different symbols represent different bands.}
\end{center}
\end{figure*}

\begin{table}
\begin{center}
\caption{Photometric solutions for GSC 03553-00845}
\begin{tabular}{lcc}
\hline
Parameters & Photometric elements & Errors  \\

\hline
    $ g_1=g_2$ &  0.32 & Assumed \\
     $A_1=A_2$ &  0.5 & Assumed\\
     $x_{1bol},$ $x_{2bol}$& 0.644, 0.647& Assumed\\

     $y_{1bol},$ $y_{2bol}$& 0.231, 0.221& Assumed\\

     $x_{1B},$ $x_{2B}$& 0.817, 0.829& Assumed\\

     $y_{1B},$ $y_{2B}$& 0.215, 0.185& Assumed\\

     $x_{1V},$ $x_{2V}$& 0.728, 0.745& Assumed\\

     $y_{1V},$ $y_{2V}$& 0.269, 0.256& Assumed\\

     $x_{1R},$ $x_{2R}$& 0.635, 0.653& Assumed\\

     $y_{1R},$ $y_{2R}$& 0.276, 0.246& Assumed\\

    $ T_1(K) $&  6250&     Assumed    \\

  $T_2(K) $&      6044 & $ \pm10$\\

    $i(deg)$ &     73.143 & $ \pm0.150$\\

  $q(M_2/M_1) $& 2.904  & $ \pm0.012$   \\

  $ \Omega_{in} $ & 6.4883 &  Assumed \\

  $ \Omega_{out} $ & 5.8708& Assumed  \\

$L_{1B}/L_B$&      0.3242 & $ \pm0.0015$\\
$L_{1V}/L_V$&      0.3124 & $ \pm0.0012$\\
$L_{1V}/L_R$&      0.3061 & $ \pm0.0010$\\

$\Omega_1$=$\Omega_2$ & 6.3061 & $ \pm0.0188$\\

  $r_1(pole)$ &   0.2846 &$ \pm 0.0017$\\

  $r_1(side)$ &   0.2985 & $ \pm0.0021$\\

  $r_1(back)$ &     0.3425 & $ \pm0.0039$\\

  $r_2(pole)$ &   0.4574 &$ \pm 0.0013$\\

  $r_2(side)$ &   0.4933 & $ \pm0.0019$\\

  $r_2(back)$ &   0.5239 & $ \pm0.0025$\\

  $f$    &   29.5\% & $\pm3.0\%$ \\

\hline
\end{tabular}
\end{center}
\end{table}

\section{Discussion and conclusions}
In this paper, we presented $BVR$ light curves of a newly discovered W UMa binary GSC 03553-00845. The orbital period of GSC 03553-00845 was determined as $P=0.435470$ by using the Jurkevich method.
Photometric analysis was carried out based on the observed light curves.
It is shown that GSC 03553-00845 is a W-subtype of a W UMa type binary ($q=2.904$), where the primary component star is the hotter and less massive one. The solution also reveals that GSC 03553-00845 is an overcontact binary system by a contact degree of $f=29.5\%$ with a small temperature difference between the components ($\Delta T=206$ K) indicating a good thermal contact between the components. The 29.5\% fill-out factor is somewhat unusual for a W-subtype W UMa binary, but it is certainly not enough to eliminate the validity of our model. For an unambiguous solution, a radial velocity curve is needed and it should be modeled simultaneously with the light curves.

As shown in Figure 2, the Max. I is higher than the Max. II., the V band light curve of GSC 03553-00845 based on all data shows obvious O'Connell effect. But the light curves of GSC 03553-00845 observed on April 29, 30 and May 18, 2014 are almost symmetrical. No spots are modeled, but strong magnetic activity is likely. This is shown by the night to night differences in the photometry and the W-subtype result. The primary component is likely covered by spots thus hiding its true, hotter temperature. Future observers will likely see the effects of spots in their light curves. Orbital period investigation plays an important role in the evolution and searching for tertiary companion for W UMa type binaries.
Further observations should be carried out to determine the orbital evolution of the binary and the possibility of third components and planets.

\acknowledgments
This work is partly supported by the National Natural Science Foundation of China (Nos.
11203016, 11333002, 10778619, 10778701), and by the
Natural Science Foundation of Shandong Province (No. ZR2012AQ008) and by the Open Research Program of Key Laboratory for the Structure and Evolution of Celestial Objects (No. OP201303). We highly appreciate the useful comments of the referee, these help us improved the paper greatly.

\end{document}